\documentclass[useAMS,usenatbib,letterpaper]{mn2e}

\usepackage{natbib}
\usepackage{bm}
\usepackage{times}
\usepackage{graphicx}
\usepackage{color}
\usepackage{amssymb}
\usepackage{fancyvrb}
\usepackage{url}
\usepackage{verbatim}

\renewcommand{\cite}{\citep}
\newcommand{\HI}{H$\,${\sc i}}
\newcommand{\rHI}{{\rm HI}}
\newcommand{\Ob}{\Omega_{\rHI} b_{\rHI}}
\newcommand{\fhr}{\rm deep}
\newcommand{\ohr}{\rm wide}
\newcommand{\ud}{\,\mathrm{d}}
\newcommand{\mat}{\mathbfss}
\newcommand{\vect}{\bmath}
\newcommand{\trans}{{\rm T}}
\newcommand{\kunit}{h\,{\rm Mpc}^{-1}}

\VerbatimFootnotes

\title[21\,cm auto-power]{Determination of $z \sim 0.8$ neutral hydrogen fluctuations using the 21\,cm intensity mapping autocorrelation}

\author[E.~R.~Switzer, K.~W.~Masui, et al.]{\parbox{\textwidth}{E.~R.~Switzer$^{1}$\thanks{E-mail:~\texttt{eswitzer@cita.utoronto.ca}},
K.~W.~Masui$^{1,2}$\thanks{E-mail:~\texttt{kiyo@cita.utoronto.ca}},
K.~Bandura$^{3}$, L.-M. Calin$^{1}$, T.-C.~Chang$^{4}$, \\
X.-L.~Chen$^{5,6}$, Y.-C.~Li$^{5}$, Y.-W.~Liao$^{4}$, A.~Natarajan$^{7}$,
U.-L.~Pen$^{1}$, J.~B.~Peterson$^{7}$, J.~R.~Shaw$^{1}$, T.~C.~Voytek$^{7}$} 
\vspace{0.4cm}\\
\parbox{\textwidth}{
$^{1}$Canadian Institute for Theoretical Astrophysics, University of Toronto, 60 St George St, Toronto, Ontario M5S 3H8, Canada \\
$^{2}$Department of Physics, University of Toronto, 60 St George St, Toronto, Ontario M5S 1A7, Canada \\
$^{3}$Department of Physics, McGill University, 3600 Rue University, Montreal, Quebec H3A 2T8, Canada \\
$^{4}$Academia Sinica Institute of Astronomy and Astrophysics, PO Box 23-141, Taipei 10617, Taiwan \\
$^{5}$National Astronomical Observatories, Chinese Academy of Science, 20A Datun Road, Beijing 100012, China \\
$^{6}$Center of High Energy Physics, Peking University, Beijing 100871, China \\
$^{7}$McWilliams Center for Cosmology, Department of Physics, Carnegie Mellon
University, 5000 Forbes Ave, Pittsburgh PA 15213, USA}}

\begin{document}

\date{April 12, 2013} 

\pagerange{\pageref{firstpage}--\pageref{lastpage}} \pubyear{2013}

\maketitle

\begin{abstract} The large-scale distribution of neutral hydrogen in the
Universe will be luminous through its $21$\,cm emission. Here, for the first
time, we use the auto-power spectrum of $21$\,cm intensity fluctuations to
constrain neutral hydrogen fluctuations at $z \sim 0.8$. Our data were acquired
with the Green Bank Telescope and span the redshift range $0.6 < z < 1$ over
two fields totalling $\approx 41\,{\rm deg}^2$~and 190\,h of radio integration time.
The dominant synchrotron foregrounds exceed the signal by $\sim 10^3$, but have
fewer degrees of freedom and can be removed efficiently. Even in the presence
of residual foregrounds, the auto-power can still be interpreted as an upper
bound on the 21\,cm signal. Our previous measurements of the cross-correlation
of $21$\,cm intensity and the WiggleZ galaxy survey provide a lower bound.
Through a Bayesian treatment of signal and foregrounds, we can combine both
fields in auto- and cross-power into a measurement of $\Ob =
[0.62^{+0.23}_{-0.15}] \times 10^{-3}$ at 68\% confidence with 9\% systematic
calibration uncertainty, where $\Omega_{\rHI}$ is the neutral hydrogen (\HI)
fraction and $b_{\rHI}$ is the \HI\ bias parameter. We describe observational
challenges with the present data set and plans to overcome them.
\end{abstract}

\label{firstpage}


\section{Introduction}

There is substantial interest in the viability of cosmological structure surveys that map the intensity of 21\,cm emission from neutral hydrogen. Such surveys could be used to study large-scale structure (LSS) at intermediate redshifts, or to study the epoch of reionization at high redshift.  Surveys of 21\,cm intensity have the potential to be very efficient since the resolution of the instrument can be matched to the large scales of cosmological interest \citep{2008PhRvL.100i1303C, 2008PhRvL.100p1301L, 2010ApJ...721..164S, 2012A&A...540A.129A}. Several experiments, including BAOBAB \citep{2013AJ....145...65P}, BAORadio \citep{2012CRPhy..13...46A}, BINGO \citep{2012arXiv1209.1041B}, CHIME\footnote{\url{http://chime.phas.ubc.ca/}}  and TianLai \citep{2012IJMPS..12..256C} propose to conduct redshift surveys from $z\sim0.5$ to $2.5$ using this method.

The principal challenges for 21\,cm experiments are astronomical foregrounds and terrestrial radio frequency interference (RFI). Extragalactic sources and the Milky Way produce synchrotron emission that is three orders of magnitude brighter than the 21\,cm signal.  However, the physical process of synchrotron emission is known to produce spectrally smooth radiation, occupying few degrees of freedom along each line of sight. In the absence of instrumental effects, these degrees of freedom are thought to be separable from the signal \citep{2011PhRvD..83j3006L, 2012MNRAS.419.3491L, 2013arXiv1302.0327S}. RFI can be minimized through site location, sidelobe control and band selection. In the Green Bank Telescope (GBT) data analysed here, RFI is not found to be a significant challenge or limiting factor.

Subtraction of synchrotron emission has proven to be challenging in practice. Instrumental effects such as passband calibration and polarization leakage couple bright foregrounds into new degrees of freedom that need to be removed from each line of sight to reach the level of the $21$\,cm signal. The spectral functions describing these systematics cannot all be modelled in advance, so we take an empirical approach to foreground removal by estimating dominant modes from the covariance of the map itself. This method requires more caution because it also removes cosmological signal, which must be accounted for.

Large-scale neutral hydrogen fluctuations above redshift ${z=0.1}$ have been unambiguously detected only in cross-correlation with existing surveys of optically selected galaxies \citep{2009MNRAS.399.1447L, 2010Natur.466..463C, 2013ApJ...763L..20M}. Here, residual 21\,cm foregrounds boost the errors, but do not correlate with the optical galaxies. The density fluctuations traced by survey galaxies may not correlate perfectly with the emission of neutral hydrogen, so their cross-correlation can be interpreted as a lower limit on the fluctuation power of 21\,cm emission.

Several efforts have used the 21\,cm line to place upper bounds on the reionization era \citep{1986MNRAS.218..577B, 2010Natur.468..796B, 2013arXiv1301.5906P, 2013arXiv1301.7099P} and $z\sim 3$ (see, e.g., \citet{1990JApA...11..221S, 1992A&A...256..331W}) without the need to cross-correlate with an external data set. This is the first work to describe similar bounds for $z\sim 0.8$, using two fields totalling $\approx 41\,{\rm deg}^2$~and 190\,h of radio integration time with the GBT. Unlike the bounds from reionization, for which there is currently no cross-correlation, we are able to combine the auto- and cross-powers in a novel way, making a Bayesian inference of the amplitude of neutral hydrogen fluctuations, parametrized by $\Omega_{\rm HI} b_{\rm HI}$. Throughout, we use cosmological parameters from \citet{2009ApJS..180..330K}.

\section{Observations and Analysis}
\label{ss:surveymaps}

The analysis here is based on the same observations used for the cross-correlation measurement in \citet{2013ApJ...763L..20M}. We flag RFI in the data, calculate 3D intensity map volumes, clean foreground contamination, and estimate the power spectrum. Here, we will summarize essential aspects of the observations and analysis in \citet{2013ApJ...763L..20M}, and describe the auto-power analysis in more detail.

Observations were conducted with the $680-920$\,MHz prime-focus receiver at the GBT, sampled from $700$\,MHz ($z = 1$) to $900$\,MHz ($z = 0.58$) in 256 uniform spectral bins. The analysis here uses a 105\,h integration of a $4\fdg5\times2\fdg4$ $15$\,h `deep' field centred on $14^\mathrm{h}31^\mathrm{m}28\fs5$ right ascension, $2^\circ0'$ declination and an 84\,h integration on a $7\fdg0 \times 4\fdg3$ 1\,h `wide' field centred on $0^\mathrm{h}52^\mathrm{m}0^\mathrm{s}$ right ascension, $2^\circ9'$ declination.

The beam full width at half-maximum at $700$\,MHz is $0\fdg314$ and at $900$\,MHz it is $0\fdg250$.  At band-centre, the beam width corresponds to a comoving length of $9.6\,h^{-1}\,{\rm Mpc}$. Both fields have nearly complete angular overlap and good redshift coverage with the WiggleZ Dark Energy Survey \citep{2010MNRAS.401.1429D}. Our absolute calibration is determined from radio point sources and is accurate to $9\%$ \citep{2013ApJ...763L..20M}. For clarity, this remains as a separately quoted systematic error throughout, and plotted posterior distributions are based on statistical errors only.

\subsection{Foreground cleaning}
\label{ss:foregrounds}

In this section, we develop the map cleaning formalism and discuss its connection to survey strategy. We begin by packing the three-dimensional map into an $N_\nu \times N_\theta$ matrix $\mat{M}$ by unwrapping the $N_\theta$ RA, Dec pointings. For the moment, we ignore thermal noise in the map. The empirical $\nu-\nu'$ covariance of the map is ${\mat{C} = \mat{M} \mat{M}^\trans/N_\theta}$, and it contains both foregrounds and 21\,cm signal. This can be factored as $\mat{C} = \mat{U} \mathbf{\Lambda} \mat{U}^\trans$, where $\mathbf{\Lambda}$ is a diagonal matrix and is sorted in descending value. From each line of sight, we can then subtract a subset of the modes $\mat{U}$ that describe the largest components of the variance through the operation $(1-\mat{U} \mat{S} \mat{U}^\trans) \mat{M}$, where $\mat{S}$ is a selection matrix with $1$ along the diagonal for modes to be removed and $0$ elsewhere. 

In reality, $\mat{M}$ also contains thermal noise. To minimize its influence on our foreground mode determination, we find the noise-inverse-weighted cross-variance of two submaps from the full season of observing. Here, $\mat{C}_{AB} = (\mat{W}_A \circ \mat{M}_A)(\mat{W}_B \circ \mat{M}_B)^\trans/N_\theta$, where $A$ and $B$ denote subseason maps, $\mat{W}_A$ is the noise-inverse-variance weight per pixel of map $A$ (neglecting correlations) and $\circ$ is the element-wise matrix product. $\mat{C}_{AB}$ is no longer symmetric, and we take its singular value decomposition (SVD) instead, using the left and right singular vectors to clean maps $A$ and $B$, respectively. The weights are calculated in the noise model developed in the map-maker, but roughly track the map's integration depth and weigh against RFI. The weight is nearly separable into angle (through integration time) and frequency [through $T_{\rm sys}(\nu)$], but we average to make it formally separable and so rank-1, so that it does not increase the map rank. The weighted removal for map $A$ becomes $(1/\mat{W}_A) \circ (1-\mat{U}_A \mat{S} \mat{U}^\trans_A) \mat{W}_A \circ \mat{M}_A$, where $1/\mat{W}_A$ is the element-wise reciprocal.

Our empirical approach to foreground removal is limited by the amount of information in the maps. The fundamental limitation here surprisingly is not from the number of degrees of freedom along the line of sight, but is instead the number of independent angular resolution elements in the map \citep{Nityananda10}. To see why this is the case, notice that in the absence of noise, our cleaning algorithm is equivalent to taking the SVD of the map directly: ${\mat{M} = \mat{U} \mathbf{\Sigma} \mat{V}^\trans}$ and thus $\mat{C} \propto \mat{M} \mat{M}^\trans = \mat{U} \mathbf{\Sigma}^2 \mat{U}^\trans$, with the same set of frequency modes $\mat{U}$ appearing in both decompositions.  The rank of $\mat{C}$ coincides with the rank of $\mat{M}$ and is limited by the number of either angular or frequency degrees of freedom.

Assuming that the foreground modes all have comparable spurious overlap with the signal, one arrives at a transfer function rule of thumb $T =P_{\rm sig.\,out} / P_{\rm sig.\,in} \sim [(1 - N_{\rm m}/N_\nu)(1 - N_{\rm m}/N_{\rm res})]^2$, where $N_{\rm m}$ is the number of modes removed, $N_\nu = 256$ is the number of frequency channels and $N_{\rm res}$ is the number of resolution elements (roughly the survey area divided by the beam solid angle). A limited number of resolution elements can greatly reduce the efficacy of the foreground cleaning at the expense of signal.

The noise-weighted effective areas of the wide and deep fields are $\sim 8$ and $\sim 3\,{\rm deg}^2$, giving roughly 70 and 30 independent resolution elements at the largest beam size. The rank of $\mat{C}$ is then less than the number of available spectral bins in both cases. To recover a factor of roughly 2 to 3 in the number of resolution elements in the weighted $\nu-\nu'$ covariance, all pointings with weights above the median are re-weighted equally.

The optimal number of modes to remove coincides with the most stringent upper bound from the auto-power spectrum. For too few modes removed, the bound is limited by residual foregrounds, and for too many modes removed, it is limited by signal loss and increasing error bars. The \ohr\ field has a clear minimum from 20 to 30 modes, and we remove $30$. This optimum concerns an ensemble average of surveys, and a particular treatment in one survey may scatter low. Hence, while the \fhr\ field has a minimum at 15, we conservatively remove 10.

\subsection{Instrumental systematics}
\label{ss:systematics}

The physical mechanism of synchrotron radiation suggests that it is described by a handful of smooth modes along each line of sight \citep{2012MNRAS.419.3491L}. Instrumental response to bright foregrounds, however, can convert these into new degrees of freedom. An imperfect and time-dependent passband calibration will cause intrinsically spectrally smooth foregrounds to occupy multiple modes in our maps with non-trivial spectral structure. We control this using a pulsed electronic calibrator, averaged for each scan.

We believe that the most challenging spectral structure from foregrounds is caused by leakage of polarization into intensity. Here, each Mueller matrix element has a characteristic beam on the sky, dependent on offset from the boresight and frequency. The spectral structure converts spectrally smooth polarization into new degrees of freedom. Faraday rotation of the polarization introduces further spectral degrees of freedom. 

The leakage beam is optical in origin, mixes $\sim 10\%$ of polarization to intensity, is antisymmetric about the boresight to a good approximation and is slightly broader than the primary beam. In addition, the frequency dependence of the pure Stokes I beam mixes spatial into spectral structure. We mitigate both of these terms by convolving to a common resolution corresponding to $1.4$ times the beam size at $700$\,MHz (the largest beam). This convolution is based on a frequency-dependent beam model from source scans. Such a convolution is viable because GBT has roughly twice the resolution needed to map LSS in the linear regime.  However, this convolution reduces the number of independent resolution elements in the map by a factor of 2, increasing the challenges discussed in Section~\ref{ss:foregrounds}.

The present results are limited largely by the area of the regions and our understanding of the instrument. With a factor of roughly 10 more than the present area, the resolution could be degraded at less expense to the signal. This requires significant telescope time because the area must also be covered to roughly the same depth as our present fields.  It would however provide a significant boost in overall sensitivity for scientific goals such as measurement of the redshift-space distortions. In addition, we are investigating mapmaking that would unmix polarization using the Mueller matrix of beams, as determined from source scans.

\subsection{Power spectrum estimation}
\label{s:pse}

Our starting point for power spectral estimation is the optimal quadratic estimator described in \citet{2011PhRvD..83j3006L}. To avoid the thermal noise bias, we only consider cross-powers between four subseason maps \citep{2005MNRAS.358..833T}, labelled here as $\{ A, B, C, D\}$. Thermal noise is uncorrelated between these sections, which we have chosen to have similar integration depth and coverage. The foreground modes are determined separately for each side of the pair using the SVD of Section~\ref{ss:foregrounds}. Up to a normalization, the resulting estimator for the pair of submaps $A$ and $B$ is $\hat P(\vect{k}_i)_{A\times B} \propto (\mat{w}_A \mathbf{\Pi}_A \vect{m}_A)^\trans \mat{Q}_i \mat{w}_B \mathbf{\Pi}_B \vect{m}_B$. Here, we have unwrapped the map matrix $\mat{M}_A$ into a one-dimensional map vector $\vect{m}_A$ and written the foreground cleaning projection $(1/\mat{W}_A) \circ (1-\mat{U}_A \mat{S} \mat{U}^\trans_A) \mat{W}_A \circ \mat{M}_A$ as $\mathbf{\Pi}_A \vect{m}_A$. The weighted mean of each frequency slice of the map is also subtracted. The map weight $\mat{w}_A$ is the matrix $\mat{W}_A$ used in the SVD, but unwrapped, and along the diagonal. Procedurally, the estimator amounts to weighting both foreground-cleaned maps, taking the Fourier transform, and then summing the three-dimensional cross-pairs to find power in annuli in two-dimensional $k$-space, $\vect{k}_i =\{ k_{\perp, i}, k_{\parallel,i} \}$. The Fourier transform and binning are performed by $\mat{Q}_i$ here. We calculate six such crossed pairs from the four-way subseason split of the data, and let the average over these be the estimated power $\hat P(\vect{k}_i)$.

We calculate transfer functions to describe signal lost in the foreground cleaning and through the finite instrumental resolution. These are functions of $k_{\perp}$ and $k_{\parallel}$. The beam transfer function is estimated using Gaussian 21\,cm signal simulations that have been convolved to the effective, frequency-independent beam described in Sec~\ref{ss:systematics}. The foreground cleaning transfer function can be efficiently estimated through Monte Carlo simulations as
\begin{equation}
\label{eqn:trans}
T(\vect{k}_i) = \left \langle \frac{[\mat{w}_A \mathbf{\Pi}_{A+s} (\vect{m}_A + \vect{m}_{\rm s}) - \mat{w}_A \mathbf{\Pi}_{A} \vect{m}_A]^\trans \mat{Q}_i \vect{m}_{\rm s}}{(\mat{w}_A \vect{m}_{\rm s})^\trans \mat{Q}_i \vect{m}_{\rm s}} \right \rangle^2,
\end{equation}
where the $A+s$ subscript denotes the fact that the foreground cleaning modes have been estimated from a $\nu-\nu'$ covariance that has added 21\,cm simulation signal, $\vect{m}_{\rm s}$. This quantity is squared because cleaning is applied to both sides of the quadratic estimator of the power spectrum. The limited number of angular resolution elements (Section~\ref{ss:foregrounds}) results in an anticorrelation of the cleaned foregrounds with the signal itself, represented by the term $(\mat{w}_A \mathbf{\Pi}_{A+s} \vect{m}_A)^\trans \mat{Q}_i \vect{m}_{\rm s}$. To reduce the noise of the simulation cross-power, note that we subtract $\mat{w}_A \mathbf{\Pi}_{A} \vect{m}_A$ in the numerator. Finally, we find the weighted average of these across the four-way split of maps. We find that $300$ signal simulations are sufficient to estimate the transfer function. 

After compensating for lost signal using transfer functions for the beam and foreground cleaning, we bin the two-dimensional powers on to one-dimensional band-powers. We weight bins by their two-dimensional Gaussian inverse noise variance $\propto N(\vect{k}_i) T(\vect{k}_i)^2 /P_{\rm auto}(\vect{k}_i)^2$, where $P_{\rm auto}(\vect{k}_i)$ is the average of $\{ P_{A \times A}, P_{B \times B}, P_{C \times C}, P_{D \times D} \}$ (pairs which contain the thermal noise bias) and $N(\vect{k}_i)$ is the number of three-dimensional $k$ modes that enter a two-dimensional bin $\vect{k}_i$. In addition to the Gaussian noise weights, we impose two additional cuts in the two-dimensional $k$-power. For $k_\parallel < 0.035\,\kunit$, $k_\perp < 0.08\,\kunit$ for the \fhr\ field and $k_\perp < 0.04\,\kunit$ for the \ohr\ field, there are few harmonics in the volume, resulting in strips in the two-dimensional power spectrum where the errors are poorly estimated and strongly correlated. For $k_\perp >0.3\,\kunit$, the instrumental resolution produces significant signal loss, so this is also truncated. 

Foregrounds in the input maps and the 21\,cm signal itself are non-Gaussian, but after cleaning, the thermal noise dominates both contributions in an individual map, and Gaussian errors (see, e.g., \citet{2011ApJ...729...62D}) provide a reasonable approximation. These take as input the auto-power measurement itself (for sample variance) and $P_{A \times A}$ terms that represent the thermal noise. Sample variance is significant only in the \fhr\ field in the lower 1/3 of the reported wavenumbers. Gaussian errors agree with the standard deviation of the six crossed pairs that enter the spectral estimation in the regime where sample variance is negligible.

The finite survey size and weights result in correlations between adjacent $k$-bins. We apodize in the frequency direction using a Blackman window and in the angular direction using the map weight itself (which falls off at the edges due to scan coverage). The bin-bin correlations are estimated using 3000 signal plus thermal noise simulations assuming $T_{\rm sys} = 25$\,K. To construct a full covariance model, these are then recalibrated by the outer product of the Gaussian error amplitudes for the data relative to the thermal noise simulation errors. 

The Bayesian method developed in the next section assumes that adjacent bins are uncorrelated. To achieve this, we take the matrix square root of the inverse of our covariance model matrix and normalize its rows to sum to one. This provides a set of functions which decorrelates \cite{2000MNRAS.312..285H} the pre-whitened power spectrum and boosts the errors. At large scales ($k=0.1\,\kunit$) where these effects are relevant, decorrelation and sample variance increase the errors by a factor of 1.5 in the \ohr\ field and 4 in the \fhr\ field.


\begin{figure}
\centerline{\includegraphics[scale=0.43]{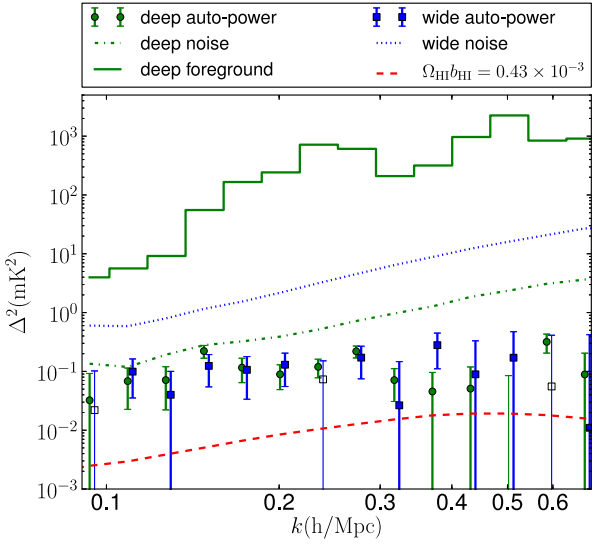}}
\caption{\label{f:ps}
Temperature scales in our 21\,cm intensity mapping survey. The top curve is the power spectrum of the input \fhr\ field with no cleaning applied (the \ohr\ field is similar). Throughout, the \fhr\ field results are green and the \ohr\ field results are blue. The dotted and dash-dotted lines show thermal noise in the maps. The power spectra avoid noise bias by crossing two maps made with separate data sets. The points below show the power spectrum of the \fhr\ and \ohr\ fields after the foreground cleaning described in Section~\ref{ss:foregrounds}. Individual modes in the map are dominated by thermal noise rather than residual foregrounds or signal. Errors are the thermal noise power divided by the number of modes in the $k$-bin, plus sample variance. The negative values are shown with thin lines and hollow markers. The red dashed line shows the 21\,cm signal expected from the amplitude of the cross-power with the WiggleZ survey (for $r=1$) and based on simulations processed by the same pipeline.}
\end{figure}

\section{Results}
\label{sec:results}

The auto-power spectra presented in Fig.~\ref{f:ps} will be biased by an unknown positive amplitude from residual foreground contamination. These data can then be interpreted as an upper bound on the neutral hydrogen fluctuation amplitude, $\Omega_{\rHI}b_{\rHI}$. In addition, we have also measured the cross-correlation with the WiggleZ galaxy survey \citep{2013ApJ...763L..20M}. This finds $\Omega_{\rHI}b_{\rHI}r=[0.43\pm0.07({\rm stat.}) \pm0.04({\rm sys.})]\times10^{-3}$, where $r$ is the WiggleZ galaxy-neutral hydrogen cross-correlation coefficient (taken here to be independent of scale). Since $|r| < 1$ by definition and is measured to be positive, the cross-correlation can be interpreted as a lower bound on $\Omega_{\rHI}b_{\rHI}$. In this section, we will develop a posterior distribution for the 21\,cm signal auto-power between these two bounds, as a function of $k$. We will then combine these into a posterior distribution on $\Omega_{\rHI}b_{\rHI}$.

The probability of our measurements given the 21\,cm signal auto-power and foreground model parameters is
\begin{equation}
p(\vect{d}_k|\boldsymbol{\theta}_k) =p(d^c|s_k,r)p(d^{\fhr}_k|s_k,f^{\fhr}_k)p(d^{\ohr}_k|s_k,f^{\ohr}_k).
\end{equation}
Here, $\vect{d}_k = \{d^c,d^{\fhr}_k,d^{\ohr}_k\}$ contains our cross-power and
\fhr\ and \ohr\ field auto-power measurements, while $\boldsymbol{\theta}_k = \{ s_k,r,f^{\fhr}_k,f^{\ohr}_k \}$ contains the 21\,cm signal auto-power, cross-correlation coefficient, and \fhr\ and \ohr\ field foreground contamination powers, respectively. The cross-power variable $d^c$ represents the constraint on $\Omega_{\rHI}b_{\rHI}r$ from both fields and the range of wavenumbers used in \citet{2013ApJ...763L..20M}. The band-powers $d^{\fhr}_k$ and $d^{\ohr}_k$ are independently distributed following decorrelation of finite-survey effects. We assume that the foregrounds are uncorrelated between $k$-bins and fields, also. This is conservative because knowledge of foreground correlations would yield a tighter constraint. We take $p(d^c|s_k,r)$ to be normally distributed with mean proportional to $r\sqrt{s_k}$, and $p(d^{\fhr}_k|s_k,f^{\fhr}_k)$ to be normally distributed with mean $s_k+f^{\fhr}_k$ and errors determined in Sec~\ref{s:pse} (and analogously for the \ohr\ field). Only the statistical uncertainty is included in the width of the distributions, as the systematic calibration uncertainty is perfectly correlated between cross- and auto-power measurements and can be applied at the end of the analysis.

\begin{figure}
\centerline{\includegraphics[scale=0.43]{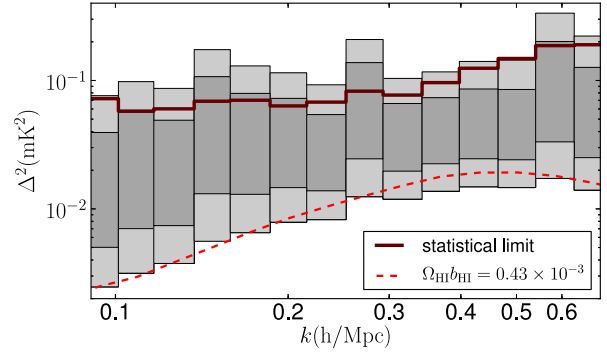}}
\caption{\label{f:allowed_signal} Comparison with the thermal noise limit. The dark and light shaded regions are the 68\% and 95\% confidence intervals of the measured 21\,cm fluctuation power from equation~(\ref{eq:bayes_signal}). The dashed line shows the expected 21\,cm signal implied by the WiggleZ cross-correlation if $r=1$.  The solid line represents the best upper $95\%$ confidence level that we could achieve given our error bars in both fields, in the absence of foreground contamination. Note that the autocorrelation measurements, which constrain the signal from above, are uncorrelated between $k$-bins, while a single global fit to the cross-power (in \citet{2013ApJ...763L..20M}) is used to constrain the signal from below. Confidence intervals do not include the systematic calibration uncertainty, which is 18\% in this space.}
\end{figure}

We apply Bayes' theorem to obtain the posterior distribution for the parameters, $p(\boldsymbol{\theta}_k| \vect{d}_k) \propto p(\vect{d}_k | \boldsymbol{\theta}_k) p(s_k)p(r)p(f^{\fhr}_k)p(f^{\ohr}_k).$ For the nuisance parameters, we adopt conservative priors. $p(f^{\fhr}_k)$ and $p(f^{\ohr}_k)$ are taken to be flat over the range $0<f_k<\infty$. Likewise, we take $p(r)$ to be constant over the range $0<r<1$, which is conservative given the theoretical bias towards $r \approx 1$. Our goal is to marginalize over these nuisance parameters to determine $s_k$. We choose the prior on $s_k$, $p(s_k)$, to be flat, which translates into a prior $p(\Ob) \propto \Ob$. The signal posterior is
\begin{equation}
p(s_k|\vect{d}_k)=\int p(s_k, r, f^{\fhr}_k, f^{\ohr}_k | \vect{d}_k) \ud r \ud f^{\fhr}_k\ud f^{\ohr}_k.
\label{eq:bayes_signal}
\end{equation}
This involves integrals of the form $\int_0^1 p(d^c|s,r)p(r)\ud{r}$ which, given the flat priors that we have adopted, can generally be written in terms of the cumulative distribution function of $p(d^c|s,r)$. Fig.~\ref{f:allowed_signal} shows the allowed signal in each spectral $k$-bin.

\begin{figure}
\centerline{\includegraphics[scale=0.43]{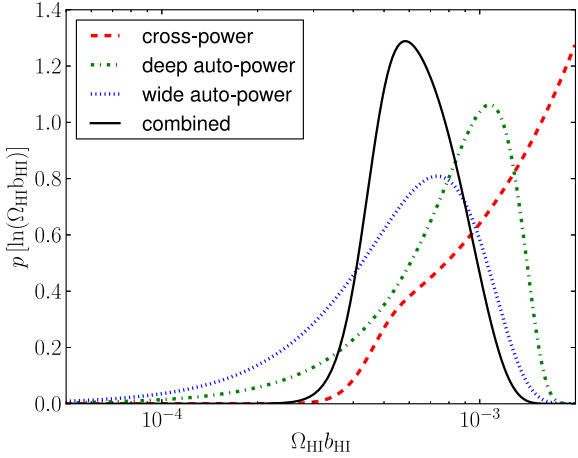}}
\caption{\label{f:likelihood} The posterior distribution for the parameter $\Omega_{\rHI} b_{\rHI}$ coming from the WiggleZ cross-power spectrum, \fhr\ field and \ohr\ field auto-powers as well as the joint likelihood from all three data sets. The priors are described in Section~\ref{sec:results}. The distributions do not include the systematic calibration uncertainty of 9\%.}
\end{figure}

Taking the analysis further, we combine band-powers into a single constraint on $\Ob$. Following \citet{2013ApJ...763L..20M}, we consider a conservative $k$-range where errors are better estimated ($k>0.12\,\kunit$ to avoid edge effects in the decorrelation operation) and before uncertainties in non-linear structure formation become significant ($k<0.3\,\kunit$). Fig.~\ref{f:likelihood} shows the resulting posterior distribution.

Our analysis yields $\Ob = [0.62^{+0.23}_{-0.15}] \times 10^{-3}$ at 68\% confidence with 9\% systematic calibration uncertainty. The range of allowed $\Ob$ is bracketed by the cross- and auto- power measurements, and is a robust statement. The peak of the posterior between these bounds is sensitive to the prior choice, and so the quoted posterior should be interpreted in the context of our prior choices here. Another reasonable signal prior is that $P(\Ob)$ is flat, which shifts the central value by $\sim 10\%$. Note that we are unable to calculate a goodness of fit to our model because each measurement is associated with a free foreground parameter which can absorb any anomalies.

\section{Discussion and Conclusions}

Through the measurement of the auto-power, we extend our previous cross-power measurement of $\Ob r$ \citep{2013ApJ...763L..20M} to a determination of $\Ob$. This is the first constraint on the amplitude of 21\,cm fluctuations at $z\sim 0.8$, and it circumvents the degeneracy with the cross-correlation $r$. The 21\,cm auto-power yields a true upper bound because it derives from the integral of the mass function. In the future, redshift distortions \citep{2008arXiv0804.1624W, 2010PhRvD..81j3527M} can be used to further break the degeneracy between $b_{\rHI}$ and $\Omega_{\rHI}$, and complement challenging {\it Hubble Space Telescope} measurements of $\Omega_{\rHI}$ \citep{2006ApJ...636..610R}. Our present survey is limited by area and sensitivity, but we have shown that foregrounds can be suppressed sufficiently, to nearly the level of the 21\,cm signal, using an empirical mode subtraction method. Future surveys exploiting the auto-power of 21\,cm fluctuations must develop statistics that are robust to the additive bias of residual foregrounds and that control instrumental systematics such as polarized beam response and passband stability.

\section*{Acknowledgements}

We would like to thank John Ford, Anish Roshi and the rest of the GBT staff for
their support; and Paul Demorest and Willem van-Straten for help with pulsar
instruments and calibration.

The National Radio Astronomy Observatory is a facility of the National Science
Foundation operated under cooperative agreement by Associated Universities,
Inc. This research is supported by NSERC Canada and CIFAR. JBP and TCV acknowledge
NSF grant AST-1009615. XLC acknowledges the Ministry of Science and Technology
Project 863 (2012AA121701), The John Templeton Foundation and NAOC Beyond the
Horizon program, and The NSFC grant 11073024. AN acknowledges financial support
from the McWilliams Center for Cosmology. Computations were performed on the
GPC supercomputer at the SciNet HPC Consortium. SciNet is funded by the Canada
Foundation for Innovation.

\bibliographystyle{mn2e}
\bibliography{main}

\begin{thebibliography}{29}
\expandafter\ifx\csname natexlab\endcsname\relax\def\natexlab#1{#1}\fi

\bibitem[{{Ansari} {et~al}\mbox{.}(2012{\natexlab{a}}){Ansari}, {Campagne},
  {Colom}, {Le Goff}, {Magneville}, {Martin}, {Moniez}, {Rich}, \&
  {Y{\`e}che}}]{2012A&A...540A.129A}
{Ansari} R. {et~al.}, 2012{\natexlab{a}}, A\&A, 540, A129

\bibitem[{{Ansari} {et~al}\mbox{.}(2012{\natexlab{b}}){Ansari}, {Campagne},
  {Colom}, {Magneville}, {Martin}, {Moniez}, {Rich}, \&
  {Y{\`e}che}}]{2012CRPhy..13...46A}
{Ansari} R., {Campagne} J.-E., {Colom} P., {Magneville} C., {Martin} J.-M.,
  {Moniez} M., {Rich} J., {Y{\`e}che} C., 2012{\natexlab{b}}, C. R. Phys., 13,
  46

\bibitem[{{Battye} {et~al}\mbox{.}(2012){Battye}, {Brown}, {Browne}, {Davis},
  {Dewdney}, {Dickinson}, {Heron}, {Maffei}, {Pourtsidou}, \&
  {Wilkinson}}]{2012arXiv1209.1041B}
{Battye} R.~A. {et~al.}, 2012, preprint (arXiv: 1209.1041)

\bibitem[{{Bebbington}(1986)}]{1986MNRAS.218..577B}
{Bebbington} D.~H.~O., 1986, MNRAS, 218, 577

\bibitem[{{Bowman} \& {Rogers}(2010)}]{2010Natur.468..796B}
{Bowman} J.~D., {Rogers} A.~E.~E., 2010, Nature, 468, 796

\bibitem[{{Chang} {et~al}\mbox{.}(2010){Chang}, {Pen}, {Bandura}, \&
  {Peterson}}]{2010Natur.466..463C}
{Chang} T.-C., {Pen} U.-L., {Bandura} K., {Peterson} J.~B., 2010, Nature, 466,
  463

\bibitem[{{Chang} {et~al}\mbox{.}(2008){Chang}, {Pen}, {Peterson}, \&
  {McDonald}}]{2008PhRvL.100i1303C}
{Chang} T.-C., {Pen} U.-L., {Peterson} J.~B., {McDonald} P., 2008, Phys. Rev.
  Lett., 100, 091303

\bibitem[{{Chen}(2012)}]{2012IJMPS..12..256C}
{Chen} X., 2012, Int. J. Mod. Phys., 12, 256

\bibitem[{{Das} {et~al}\mbox{.}(2011){Das}, {Marriage}, {Ade}, {Aguirre},
  {Amiri}, {Appel}, {Barrientos}, {Battistelli}, {Bond}, {Brown}, {Burger},
  {Chervenak}, {Devlin}, {Dicker}, {Bertrand Doriese}, {Dunkley}, {D{\"u}nner},
  {Essinger-Hileman}, {Fisher}, {Fowler}, {Hajian}, {Halpern}, {Hasselfield},
  {Hern{\'a}ndez-Monteagudo}, {Hilton}, {Hilton}, {Hincks}, {Hlozek},
  {Huffenberger}, {Hughes}, {Hughes}, {Infante}, {Irwin}, {Baptiste Juin},
  {Kaul}, {Klein}, {Kosowsky}, {Lau}, {Limon}, {Lin}, {Lupton}, {Marsden},
  {Martocci}, {Mauskopf}, {Menanteau}, {Moodley}, {Moseley}, {Netterfield},
  {Niemack}, {Nolta}, {Page}, {Parker}, {Partridge}, {Reid}, {Sehgal},
  {Sherwin}, {Sievers}, {Spergel}, {Staggs}, {Swetz}, {Switzer}, {Thornton},
  {Trac}, {Tucker}, {Warne}, {Wollack}, \& {Zhao}}]{2011ApJ...729...62D}
{Das} S. {et~al.}, 2011, ApJ, 729, 62

\bibitem[{{Drinkwater} {et~al}\mbox{.}(2010){Drinkwater}, {Jurek}, {Blake},
  {Woods}, {Pimbblet}, {Glazebrook}, {Sharp}, {Pracy}, {Brough}, {Colless},
  {Couch}, {Croom}, {Davis}, {Forbes}, {Forster}, {Gilbank}, {Gladders},
  {Jelliffe}, {Jones}, {Li}, {Madore}, {Martin}, {Poole}, {Small}, {Wisnioski},
  {Wyder}, \& {Yee}}]{2010MNRAS.401.1429D}
{Drinkwater} M.~J. {et~al.}, 2010, MNRAS, 401, 1429

\bibitem[{{Hamilton} \& {Tegmark}(2000)}]{2000MNRAS.312..285H}
{Hamilton} A.~J.~S., {Tegmark} M., 2000, MNRAS, 312, 285

\bibitem[{{Komatsu} {et~al}\mbox{.}(2009){Komatsu}, {Dunkley}, {Nolta},
  {Bennett}, {Gold}, {Hinshaw}, {Jarosik}, {Larson}, {Limon}, {Page},
  {Spergel}, {Halpern}, {Hill}, {Kogut}, {Meyer}, {Tucker}, {Weiland},
  {Wollack}, \& {Wright}}]{2009ApJS..180..330K}
{Komatsu} E. {et~al.}, 2009, ApJS, 180, 330

\bibitem[{{Lah} {et~al}\mbox{.}(2009){Lah}, {Pracy}, {Chengalur}, {Briggs},
  {Colless}, {de Propris}, {Ferris}, {Schmidt}, \&
  {Tucker}}]{2009MNRAS.399.1447L}
{Lah} P. {et~al.}, 2009, MNRAS, 399, 1447

\bibitem[{{Liu} \& {Tegmark}(2011)}]{2011PhRvD..83j3006L}
{Liu} A., {Tegmark} M., 2011, Phys. Rev. D, 83, 103006

\bibitem[{{Liu} \& {Tegmark}(2012)}]{2012MNRAS.419.3491L}
{Liu} A., {Tegmark} M., 2012, MNRAS, 419, 3491

\bibitem[{{Loeb} \& {Wyithe}(2008)}]{2008PhRvL.100p1301L}
{Loeb} A., {Wyithe} J.~S.~B., 2008, Phys. Rev. Lett., 100, 161301

\bibitem[{{Masui}, {McDonald} \& {Pen}(2010){Masui}, {McDonald}, \&
  {Pen}}]{2010PhRvD..81j3527M}
{Masui} K.~W., {McDonald} P., {Pen} U.-L., 2010, Phys. Rev. D, 81, 103527

\bibitem[{{Masui} {et~al}\mbox{.}(2013){Masui}, {Switzer}, {Banavar},
  {Bandura}, {Blake}, {Calin}, {Chang}, {Chen}, {Li}, {Liao}, {Natarajan},
  {Pen}, {Peterson}, {Shaw}, \& {Voytek}}]{2013ApJ...763L..20M}
{Masui} K.~W. {et~al.}, 2013, ApJL, 763, L20

\bibitem[{{Nityananda}(2010)}]{Nityananda10}
{Nityananda} R., 2010, {NCRA Technical Reports}

\bibitem[{{Paciga} {et~al}\mbox{.}(2013){Paciga}, {Albert}, {Bandura}, {Chang},
  {Gupta}, {Hirata}, {Odegova}, {Pen}, {Peterson}, {Roy}, {Shaw}, {Sigurdson},
  \& {Voytek}}]{2013arXiv1301.5906P}
{Paciga} G. {et~al.}, 2013, preprint (arXiv:1301.5906)

\bibitem[{{Pober} {et~al}\mbox{.}(2013{\natexlab{a}}){Pober}, {Parsons},
  {Aguirre}, {Ali}, {Bradley}, {Carilli}, {DeBoer}, {Dexter}, {Gugliucci},
  {Jacobs}, {MacMahon}, {Manley}, {Moore}, {Stefan}, \&
  {Walbrugh}}]{2013arXiv1301.7099P}
{Pober} J.~C. {et~al.}, 2013{\natexlab{a}}, preprint (arXiv:1301.7099)

\bibitem[{{Pober} {et~al}\mbox{.}(2013{\natexlab{b}}){Pober}, {Parsons},
  {DeBoer}, {McDonald}, {McQuinn}, {Aguirre}, {Ali}, {Bradley}, {Chang}, \&
  {Morales}}]{2013AJ....145...65P}
{Pober} J.~C. {et~al.}, 2013{\natexlab{b}}, AJ, 145, 65

\bibitem[{{Rao}, {Turnshek} \& {Nestor}(2006){Rao}, {Turnshek}, \&
  {Nestor}}]{2006ApJ...636..610R}
{Rao} S.~M., {Turnshek} D.~A., {Nestor} D.~B., 2006, ApJ, 636, 610

\bibitem[{{Seo} {et~al}\mbox{.}(2010){Seo}, {Dodelson}, {Marriner}, {Mcginnis},
  {Stebbins}, {Stoughton}, \& {Vallinotto}}]{2010ApJ...721..164S}
{Seo} H.-J., {Dodelson} S., {Marriner} J., {Mcginnis} D., {Stebbins} A.,
  {Stoughton} C., {Vallinotto} A., 2010, ApJ, 721, 164

\bibitem[{{Shaw} {et~al}\mbox{.}(2013){Shaw}, {Sigurdson}, {Pen}, {Stebbins},
  \& {Sitwell}}]{2013arXiv1302.0327S}
{Shaw} J.~R., {Sigurdson} K., {Pen} U.-L., {Stebbins} A., {Sitwell} M., 2013,
  preprint (arXiv: 1302.0327)

\bibitem[{{Subrahmanyan} \& {Anantharamaiah}(1990)}]{1990JApA...11..221S}
{Subrahmanyan} R., {Anantharamaiah} K.~R., 1990, JA\& A, 11, 221

\bibitem[{{Tristram} {et~al}\mbox{.}(2005){Tristram},
  {Mac{\'{\i}}as-P{\'e}rez}, {Renault}, \& {Santos}}]{2005MNRAS.358..833T}
{Tristram} M., {Mac{\'{\i}}as-P{\'e}rez} J.~F., {Renault} C., {Santos} D.,
  2005, MNRAS, 358, 833

\bibitem[{{Wieringa}, {de Bruyn} \& {Katgert}(1992){Wieringa}, {de Bruyn}, \&
  {Katgert}}]{1992A&A...256..331W}
{Wieringa} M.~H., {de Bruyn} A.~G., {Katgert} P., 1992, A\&A, 256, 331

\bibitem[{{Wyithe}(2008)}]{2008arXiv0804.1624W}
{Wyithe} S., 2008, preprint (arXiv: 0804.1624)

\end{thebibliography}

\end{document}